\documentclass[12pt]{article}
\usepackage{graphicx}
\usepackage{lineno}
\newcommand\pubnumber{CMS CR-2017/029}
\newcommand\pubdate{\today}

\def\Title#1{\begin{center} {\Large #1 } \end{center}}
\def\Author#1{\begin{center}{ \sc #1} \end{center}}
\def\Address#1{\begin{center}{ \it #1} \end{center}}

\newcommand\pubblock{\rightline{\begin{tabular}{l} \pubnumber\\
         \pubdate  \end{tabular}}}
\newenvironment{Abstract}{\begin{quotation}  }{\end{quotation}}
\newenvironment{Presented}{\begin{quotation} \begin{center} 
             PRESENTED AT\end{center}\bigskip 
      \begin{center}\begin{large}}{\end{large}\end{center} \end{quotation}}

\def\beq{\begin{equation}}
\def\eeq#1{\label{#1}\end{equation}}
\def\eeqn{\end{equation}}

\def\beqa{\begin{eqnarray}}
\def\eeqa#1{\label{#1}\end{eqnarray}}
\def\eeqan{\end{eqnarray}}

\let\bar=\overbar

\def\etal{{\it et al.}}

\def\Dslash{\not{\hbox{\kern-4pt $D$}}}
\def\dslash{\not{\hbox{\kern-2pt $\del$}}}

\def\msb{{\bar{\ssstyle M \kern -1pt S}}}

\newcommand*{\iab}{\mbox{ab$^{-1}$}}
\newcommand*{\ifb}{\mbox{fb$^{-1}$}}
\newcommand*{\TeV}{\ensuremath{\mathrm{Te\kern -0.1em V}}}
\newcommand*{\tit}[1]{\textit{#1}}
\newcommand*{\url}[1]{\textsc{url}:https://cds.cern.ch/record/#1}

\begin{document}
\begin{titlepage}
\pubblock

\vfill
\Title{Top Quark Physics at the High Luminosity LHC}
\vfill
\Author{ Patrizia Azzi}
\Address{INFN - Section of Padova, I-35010 Padova, ITALY}
\Author{Markus Cristinziani}
\Address{Physikalisches Institut, Universit\"{a}t Bonn, GERMANY}

\centering{on behalf of the ATLAS and CMS Collaboration}

\vfill
\begin{Abstract}
A summary of the prospects for top-quark physics at the High Luminosity LHC is given.
\end{Abstract}
\vfill
\begin{Presented}
$9^{th}$ International Workshop on Top Quark Physics\\
Olomouc, Czech Republic,  September 19--23, 2016
\end{Presented}
\vfill
\end{titlepage}
\def\thefootnote{\fnsymbol{footnote}}
\setcounter{footnote}{0}

\section{Introduction}

After Run 2 of the LHC the accelerator will undergo a significant upgrade phase
to be able to deliver to the experiments a total integrated luminosity of
$3\,\iab$ at a centre-of-mass energy of $14\,\TeV$. The ATLAS and CMS
collaborations will also upgrade their detectors\cite{ATLASDET,CMSDET} in order to profit from the
large total luminosity, while coping with the increased number of pile-up
interactions.  During this ``High Luminosity'' phase the LHC (HL-LHC) will
become effectively a top-quark factory with a total number of $\sim\!3$ 
billion top-quark pairs produced and $\sim\!1$ billion produced singly.  In
this summary we report the most important measurements that could be achieved
during the HL-LHC and the latest extrapolations on their uncertainties.
Precision top-quark physics is one of the main tools to probe for new physics
beyond the Standard Model (SM) in an indirect way, if no clear discovery is
made during Run 2 of the LHC.  Lastly we consider how top quarks produced in
decay of new exotic particles can increase the spectrum of direct searches for
phenomena beyond the SM.

\section{The high luminosity LHC programme}

The high luminosity programme is scheduled to start in 2024 with a long
shutdown, called LS3, during which the machine and the detectors will be
upgraded to function at a luminosity of $5\,(7.5) \times
10^{34}\,\mathrm{cm}^{-2} \mathrm{s}^{-1}$, corresponding to an average pile-up
of $\langle\mu\rangle = 140\,(200)$.  The plan is to collect a total integrated
luminosity of $3\,\iab$ by the end of the HL-LHC programme. Currently the
experiments are in the process of drafting TDR documents, describing the
upgraded detector that should allow to keep or enhance the Run 2 performance
for physics, even in the more complicated high pile-up environment. Both
detectors will undergo quite similar upgrades: the tracker and vertexing
detectors to cope with the larger pile-up and radiation and to allow triggering
on tracks at L1 of the trigger system; the electronics of the calorimeters; a
refurbishing and extension of the muon system and their electronics to increase
acceptance and trigger capabilities; and a new trigger and DAQ to cope with the
high rates. For the case of CMS the project includes also the replacement of
the forward calorimeter with a high-granularity option that in conjunction with
the increased tracker efficiency at large $\eta$ should improve reconstruction
of forward physics objects and help with the overall pile-up mitigation and
missing energy resolution. The possibility of a fast-timing layer is also being
considered by both experiments for the mitigation of pile-up and improvement in
event vertex reconstruction.  Several studies have been already performed
\cite{CMSTP, CMSSD, ATLASLOI, ATLASSD, CMSDPPerf, ATLASPerf},
that show how the upgraded detector maintains or extends the current 
capabilities for physics measurements. 

\section{Challenges of upgrade physics studies}

The process of evaluating performance of future detectors and physics potential
is always complicated. In the case of the HL-LHC the main technical challenges
are the simulation of the detector and large pile-up, which requires
significant computing resources on one hand, while the estimate of the
background from detector effects or from the fake contributions is by
definition difficult to estimate. Most of the systematic uncertainties that are
evaluated with data are assumed to get reduced, scaling with the larger
datasets available.  In the end the precision of the results will depend on the
progress of the theoretical calculations in 15 years from now. In order to
extract the numbers presented here some choices have been taken by the two
experiments.  Concerning the simulation of signal and background events, both
ATLAS and CMS have chosen the path of a fast simulation that uses a simplified
description of the detector and a parameterisation of the
response~\cite{Delphes}.

\section{Top-quark mass} 

Studies of the uncertainties on the top quark mass, extrapolated by the CMS
collaboration from measurements with $19.7\,\ifb$ of data at $8\,\TeV$,
have been updated~\cite{CMSDPPhys}.
Several methods are considered 
that suffer from different theoretical systematic uncertainties,
related to the definition of the top-quark mass.  
Different assumptions enter the calculation of the updated
uncertainties. It is assumed that the pile-up mitigation
techniques will be adequate to keep the effects under control, in particular
no degradation of the jet resolution is expected. It is further assumed that the loss of
efficiency due to increased thresholds will be compensated by the higher cross
section at $14\,\TeV$. The statistical uncertainty is expected to scale with the
square root of the collected integrated luminosity. This might be a conservative
assumption as the increased acceptance of the upgraded detector in the forward region
is not taken into account. The conclusion is that with $3\,\iab$ of
data all the measurement will be systematics limited and especially by
theoretical modelling uncertainties. Conventional methods, which would remain
the most precise ones, are expected to yield an ultimate relative uncertainty of
$0.1\%$. The new extrapolations are shown in Fig.\ref{fig:topmassproj} (left). 

\section{FCNC in top-quark decays}

In the framework of the SM, top-quark flavor changing neutral currents (FCNC) are
highly suppressed. Predicted branching fractions for processes like
$t\rightarrow \gamma u (c)$ and $t\rightarrow Zu(c)$ range from
$10^{-16} (10^{-14})$ to $10^{-17} (10^{-14})$. 
These values are several
orders of magnitude below the sensitivity of current and planned experiments.
However, these branching ratios are enhanced in several extensions of the SM,
and any observation of these rare transitions would be a clear signal for a new
physics effect. Two studies have been updated to evaluate the
sensitivity of the upgraded ATLAS and CMS detectors during the HL-LHC run. The
first one by ATLAS~\cite{ATLASFCNC} concerns the search for FCNC in events with
pair-produced top quarks, decaying to final states with three leptons, 
two of which compatible with a leptonic $Z$ boson decay, at least one $b$-tagged jet,
and at least one other non-tagged jet.  
This selection allows to extract limits on the process $t\rightarrow Zq$. The
sensitivity increases by factors of two to six considering different
scenarios for the systematics uncertainties, reaching values between 8 and
20$\times10^{-5}$ in the more optimistic configuration.  A second search for the
case where one of the two top quarks decays via $t\rightarrow Hq$ with 
$H\rightarrow b\bar{b}$ is performed requiring events with large jet
multiplicity and at least two $b$-tagged jets, see
Fig.~\ref{fig:ATLASFCNC} (right). For this analysis the signal extraction
is more challenging and a multivariate approach is employed, to obtain a final
sensitivity around $10^{-4}$, which is twenty times better than previous
extrapolations. The last analysis considered is the search for events where a
single top quark is produced in association with a photon via an anomalous FCNC
vertex $tq\gamma$. The event selection requires the presence of a top quark
decaying in the leptonic channel and one isolated high $E_{\mathrm{T}}$ photon, well
separated from the top decay products. Also in this case the sensitivity increases by
a factor between three and ten (depending on the assumptions on the systematic uncertainties)
obtaining limits on  $B(t\rightarrow u+\gamma)(B(t\rightarrow c+\gamma))$
of $2.7 \times 10^{-5} (2.0 \times 10^{-4})$.  It is clear that the study of FCNC
has a large potential at the HL-LHC due to the very large statistics of
top quarks anticipated. Possibly, analyses re-optimized to profit of the full
potential of the new detector would improve even more that reach. 

\begin{figure}[htb]
\centering
\includegraphics[width=0.545\textwidth]{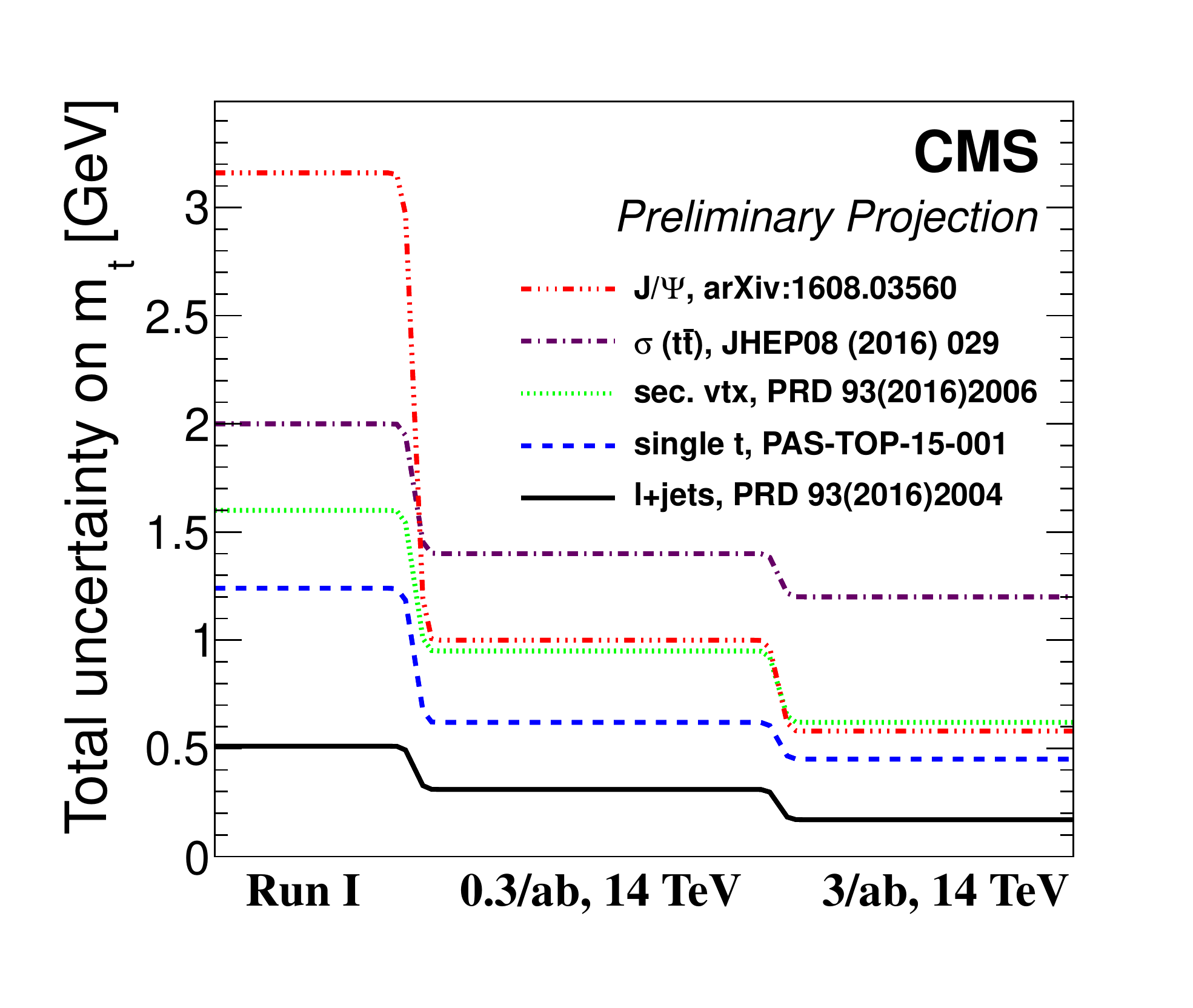}
\includegraphics[width=0.445\textwidth]{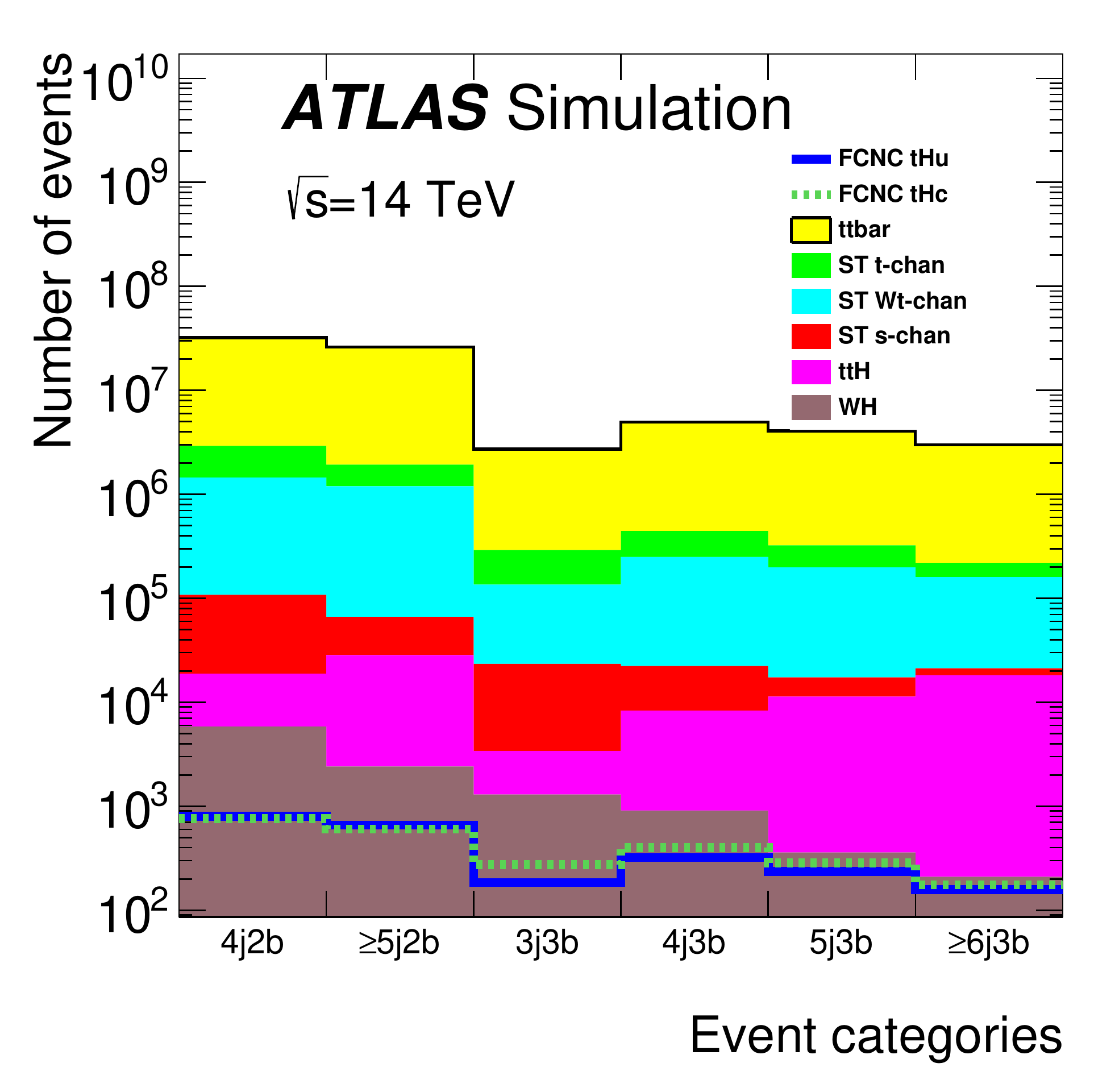}
\caption{(Left) Total top-quark mass ($m_t$) uncertainty obtained with
different measurement methods at present and their projections to the HL-LHC\cite{CMSDPPhys} .
The projections are based on measurement obtained with the CMS detector during
Run\,1 of the LHC.\newline (Right) Selected signal and background events in
the six analysis categories for the reference ATLAS detector upgrade layout\cite{ATLASFCNC}.
The signal $t \rightarrow Hu$ and $t \rightarrow Hc$ event numbers correspond to a FCNC
branching ratio of $2.67 \times 10^{-5}$. }
\label{fig:topmassproj}
\label{fig:ATLASFCNC}
\end{figure}

\section{Top quarks as portal to new physics} 

The top quark is a natural candidate, due to its large mass, to connect us to
potential new physics signatures. The analyses can profit of clean selections
that push the phase space normally examined by traditional searches. In the case
of SUSY models, for instance, measuring precisely properties of the top system,
such as spin correlations, has allowed to probe the kinematically difficult
corner for the stop pair production, close to the top-quark mass~\cite{ATLASStop}. 
In the case of a mass splitting between the gluino and the LSP close to twice the
top-quark mass, the search for a four-top final state can explore the gluino pair
production. These kind of analyses, that are being performed in 
Run~2~\cite{ATLASGluinob, ATLASGluinoLep}, will profit from a larger integrated
luminosity.  The improvement in the reconstruction of top quarks as boosted objects
and the large HL-LHC dataset will also help to push the sensitivity up to $4\,\TeV$
for the search for heavy bosons such as $Z'\rightarrow t\bar{t}$ or $W'\rightarrow
tb$~\cite{CMSDPPhys,ATLASZ}, even if a machine with larger energy would perform
better.
Finally, having a larger dataset and restricting the selection
to very boosted top-quark events will allow to improve the sensitivity to
anomalous $ttZ$ axial and vector couplings~\cite{ttZ}, $gtt$ chromo-magnetic
moments~\cite{Fuks} or top asymmetries($A_{\textrm{C}}$,$A_{\textrm{FB}}$)~\cite{Snowmass}.

\section{Summary and conclusions}

The potential opportunities for top-quark physics offered by the HL-LHC are
being explored again, profiting from the Run 2 data at $13\,\TeV$. The extrapolation
to new detectors and harsher running conditions is not trivial, and the work is
just starting. It is clear that top physics is a major item for the physics
programme of the Phase\,2: in terms of advancing the precision of the SM
measurement, in finding deviations due to new physics effects or looking for
new particles decaying to top quarks. The larger datasets of top-quark pairs
and single top will allow to perform new analyses that profit of rare
final states and extreme kinematical phase space. 

%%%%%%%%%%%%%%%%%%%%%%%%%%%%%%%%%%%%%%%%%%%%%%%%%%%%%%%%%%%%%%%%%%%%%%%%%
%%
%%   use this format to include an .eps figure into your paper
%%
%\begin{figure}[htb]
%\centering
%\includegraphics[height=1.5in]{magnet}
%\caption{Plan of the magnet used in the mesmeric studies.}
%\label{fig:magnet}
%\end{figure}
%%%%%%%%%%%%%%%%%%%%%%%%%%%%%%%%%%%%%%%%%%%%%%%%%%%%%%%%%%%%%%%%%%%%%%%%%%%

%%%%%%%%%%%%%%%%%%%%%%%%%%%%%%%%%%%%%%%%%%%%%%%%%%%%%%%%%%%%%%%%%%%%%%%%%%
%%%
%%%   use this format to include a LaTeX table  into your paper
%%%
%\begin{table}[t]
%\begin{center}
%\begin{tabular}{l|ccc}  
%Patient &  Initial level($\mu$g/cc) &  w. Magnet &  
%w. Magnet and Sound \\ \hline
% Guglielmo B.  &   0.12     &     0.10      &     0.001  \\
% Ferrando di N. &  0.15     &     0.11      &  $< 0.0005$ \\ \hline
%\end{tabular}
%\caption{Blood cyanide levels for the two patients.}
%\label{tab:blood}
%\end{center}
%\end{table}
%%%%%%%%%%%%%%%%%%%%%%%%%%%%%%%%%%%%%%%%%%%%%%%%%%%%%%%%%%%%%%%%%%%%%%%%%%%

\end{document}